\documentclass[%
 aip,
 amsmath,amssymb,
 reprint,%
]{revtex4-1}

\usepackage{graphicx}
\usepackage{dcolumn}
\usepackage{bm}

\usepackage[utf8]{inputenc}
\usepackage[T1]{fontenc}
\usepackage{mathptmx}
\usepackage{etoolbox}
\usepackage{float}
\usepackage{graphicx}
\usepackage{caption}
\usepackage{subcaption}
\usepackage{dblfloatfix}
\usepackage{amssymb}
\usepackage{amsmath}
\makeatletter
\def\@email#1#2{%
 \endgroup
 \patchcmd{\titleblock@produce}
  {\frontmatter@RRAPformat}
  {\frontmatter@RRAPformat{\produce@RRAP{*#1\href{mailto:#2}{#2}}}\frontmatter@RRAPformat}
  {}{}
}%
\makeatother
\begin{document}

\preprint{AIP/123-QED}

\title{CMA-Unfold: A Covariance Matrix Adaptation–Unfolding algorithm for stacked calorimeter detectors}

\author{G. Fauvel}
\email{gaetan.fauvel@u-bordeaux.fr}
\affiliation{University of Bordeaux, CELIA, CNRS, CEA, UMR 5107, F-33405 Talence, France}
\affiliation{ 
ELI Beamlines Facility|The Extreme Light Infrastructure ERIC
Za Radnicí 835, 252 41 Dolní Břežany, Czech Republic
}
\author{A. Arefiev}
\affiliation{Center for Energy Research, University of California San Diego, La Jolla, CA 92093, USA}

\author{M. Manuel}
\affiliation{General Atomics, San Diego, CA 92186, USA}

\author{K. Tangtartharakul}
\affiliation{Center for Energy Research, University of California San Diego, La Jolla, CA 92093, USA}

\author{S. Weber}
\affiliation{ 
ELI Beamlines Facility|The Extreme Light Infrastructure ERIC
Za Radnicí 835, 252 41 Dolní Břežany, Czech Republic
}

\author{F.P. Condamine}
\affiliation{GenF, 2 Avenue Gay Lussac, 78990 Elancourt, France}

\begin{abstract}
Stacking calorimeters also refered as bremsstrahlung cannons widely used in inertial confinement fusion and ultra-intense laser plasma experiments have become essential diagnostics for characterizing short bursts of high-energy photons and charged particles. Extracting the underlying energy spectrum from these detectors requires solving an ill-posed inverse problem, often complicated by noise, secondary particle contamination, and uncertainties in the detector response. In this work, we introduce an open-source unfolding framework (\texttt{ggfauvel/CMA-unfold}) based on the Covariance Matrix Adaptation Evolution Strategy (CMA-ES), designed to reconstruct photon spectra directly from depth–dose profiles without imposing restrictive parametric assumptions. The algorithm demonstrates high robustness, accurately recovering complex spectral shapes while tolerating percent-level deviations in individual detector layers. This approach provides a flexible and noise-resilient tool for the analysis of stacking calorimeter data, with particular relevance for bremsstrahlung diagnostics in inertial confinement fusion and high-intensity laser applications.
\end{abstract}
\date{\today}

\maketitle

\section{Introduction}
\label{sec1}

In high-energy-density and hot-plasma physics, accurate characterization of hard X-ray and gamma-ray emission from high energy electrons is essential for both fundamental studies and facility performance. Bremsstrahlung diagnostics are widely used to quantify the energy content and angular distribution of hot electrons generated in ultra-intense laser–solid interactions, where dense relativistic beams drive bright radiographic sources and secondary particle production \cite{tavana_ultra-high_2023,rhee_spectral_2016}. In Inertial Confinement Fusion (ICF), similar measurements assess how parametric instabilities, transport and preheat influence implosion performance, making hard-X-ray diagnostics key to target design and campaign optimization \cite{chaurasia_optimization_2013,ghosh_effect_2022,tommasini_development_2011}. These needs have motivated the development of dedicated spectrometers capable of operating in harsh electromagnetic and neutron environments and at high repetition rates.

A versatile class of instruments, commonly named bremsstrahlung cannons or stacking calorimeters, uses a longitudinal sequence of passive or active layers to infer the incident photon or particle spectrum from the depth–dose profile. At ICF-scale facilities, image-plate systems such as CRACC-X at the Laser Mégajoule (LMJ) \cite{noauthor_user_2024} combine filters and absorbers to obtain temperature-resolved X-ray spectra while remaining compact for target-bay deployment \cite{koester_bremsstrahlung_2021}. In the short-pulse community, analogous concepts are long established for proton and ion beams through radiochromic-film (RCF) stacks \cite{nurnberg_radiochromic_2009,kirby_radiation_2011,seimetz_spectral_2018,breschi_new_2004}, and more recently scintillator-based stacks coupled to optical readout have enabled real-time monitoring \cite{fauvel_compact_2025,istokskaia_real-time_2024}. Despite differences in implementation, these diagnostics share the same principle: harder spectra penetrate deeper, such that the recorded longitudinal profile can be inverted to retrieve the underlying energy distribution.

This inversion constitutes an ill-posed spectral-unfolding problem: each layer’s signal corresponds to the unknown spectrum convolved with a response matrix obtained from transport simulations \cite{reginatto_spectrum_2002}. Existing unfolding approaches for stacked detectors, developed for both proton and bremsstrahlung measurements, are often facility-specific, incorporate parametric assumptions, or require expert tuning of regularization and prior models \cite{breschi_new_2004,kirby_radiation_2011,craveiro_unfolding_2024}. As a result, broadly applicable and reproducible open-source tools for unfolding spectra from stacking calorimeters remain scarce, even as experimental reliance on such diagnostics increases.

At the same time, advanced optimization and machine-learning approaches have been shown to perform well on high-dimensional, noisy and non-convex inverse problems. Among derivative-free methods, the Covariance Matrix Adaptation Evolution Strategy (CMA-ES) has become a leading optimizer for continuous, ill-conditioned objective functions due to its adaptive covariance learning and robustness to noise and local minima \cite{nikolaus_hansen_cma_2016,beyer_toward_2017}. Recent studies have begun applying evolutionary strategies to automated reconstruction of electron and photon spectra, including compact gamma-ray spectrometers for laser plasma experiments \cite{fauvel_compact_2025,schmitz_automated_2022}. Yet no general, community-oriented framework employing CMA-ES for unfolding in stacking calorimeters has been documented.

This work addresses this gap by presenting an open-source unfolding framework based on CMA-ES for reconstructing particle and photon spectra from stacking calorimeter measurements. The algorithm takes the unfolding procedure as an optimization over a discretized spectrum and avoids fixed parametric assumptions, such as those related to the photon generation mechanism in the initial guess. Layer responses are computed from pre-tabulated or simulated response functions, and the discrepancy between measured and predicted depth–dose profiles defines a cost function minimized by CMA-ES under physically motivated constraints such as non-negativity and optional smoothness. Because the optimization engine is decoupled from the physics model, the method handles arbitrary geometries and materials, integrates naturally with Monte-Carlo-generated response matrices, and applies readily to both photon and charged-particle stacks.

The study presented in this article begins with a concise overview of the CMA-ES method, followed by the different weights used in the algorithm and finishing with performance tests on synthetic datasets spanning spectral shapes relevant to high-intensity-laser and ICF conditions, as well as on representative experimental measurements from bremsstrahlung cannon and stacking-calorimeter campaigns. These benchmarks collectively demonstrate the capabilities and versatility of the method.

\section{Covariance Matrix Adaptation Evolution Strategy (CMA-ES)}
\label{sec2}

\subsection{Covariance Matrix Adaptation Evolution Strategy (CMA-ES) working principle}

The Covariance Matrix Adaptation Evolution Strategy operates as a stochastic evolutionary algorithm \cite{nikolaus_hansen_cma_2016,hansen_cma-espycma_2024}, inspired by biological evolution, and iteratively enhances a population of candidate solutions based on natural selection and genetic variation principles. This method was for our case, the fastest and most reliable stochastic algorithm. To give a simple idea of its working principle, a new population is created with its corresponding energy deposition pattern. The difference between the experimental data is calculated, and then a new population is created.

The algorithm starts with an initial random population of candidate solutions, $\mathbf{x}_i, i = 1, 2, ..., \lambda$, where $\lambda$ denotes the population size. Each candidate is a vector in the $n$-dimensional real-valued solution space.

The evolutionary process starts with selecting the fittest individuals, akin to natural selection. CMA-ES uniquely uses the entire population's information to adapt its search strategy. The covariance matrix, updated each iteration, encodes variable relationships. The fitness of each candidate is evaluated using the fitness function, a function to minimize described in Section \ref{sec3} 

Subsequently, the algorithm selects the $\mu$ best solutions based on this fitness, with $\mu < \lambda$.

The updated covariance matrix influences the generation of new candidates. By sampling from a multivariate normal distribution defined by this matrix, CMA-ES balances exploration and exploitation in the solution space. Additionally, the algorithm incorporates a recombination mechanism to produce offspring from the best solutions, mirroring biological genetic recombination using the formula from the covariance matrix $\mathbf{C}$ :
\begin{equation}
\mathbf{C} \leftarrow (1 - c_{cov})\mathbf{C} + c_{cov} \frac{1}{\mu} \sum_{i=1}^{\mu} (\mathbf{x}_{i} - \mathbf{m})(\mathbf{x}_{i} - \mathbf{m})^T
\end{equation}
where $\mathbf{m}$ is the mean of the best solutions, and $c_{cov}$ is the learning rate for the covariance matrix update.

The step size $\sigma$ is crucial for determining the algorithm's progression in the solution space. Its adaptation employs a mechanism akin to the 1/5th success rule:
\begin{equation}
\sigma \leftarrow \sigma \exp\left(\frac{c_{\sigma}}{d_{\sigma}}\left(\frac{\|\mathbf{p}_{\sigma}\|}{E\|\mathcal{N}(0, \mathbf{I})\|} - 1\right)\right)
\end{equation}
where $\mathbf{p}_{\sigma}$ is the evolution path, $c_{\sigma}$ the step size learning rate, $d_{\sigma}$ the damping parameter, and $E\|\mathcal{N}(0, \mathbf{I})\|$ the expected length of a random vector from a standard normal distribution.

\begin{figure}[H]
\centering
\includegraphics[width = 0.48\textwidth]{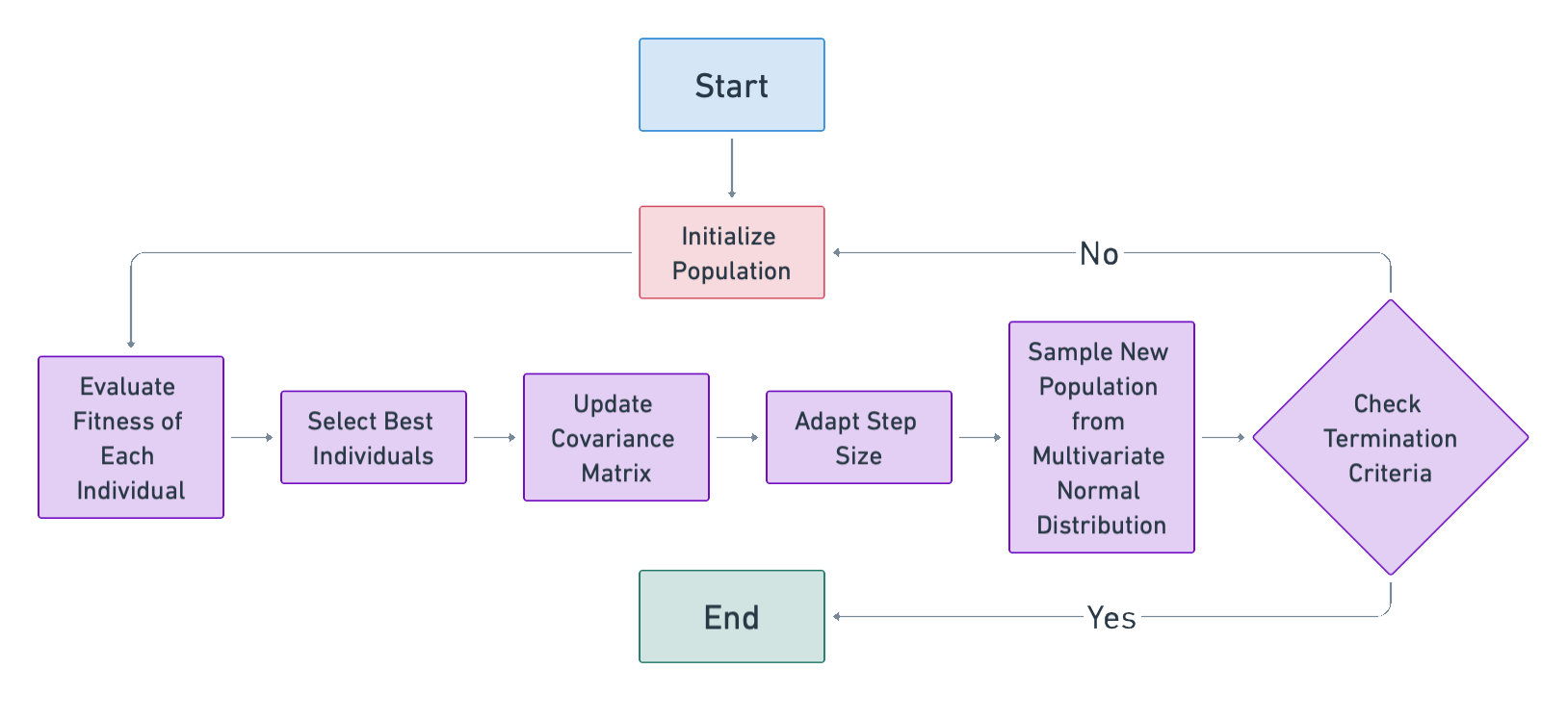}
\caption{CMA-ES principle flowing chart.}
\label{fig:CMA_ES_principle}
\end{figure}

New candidates for the next generation are sampled as follows:
\begin{equation}
\mathbf{x}_{i}' \sim \mathcal{N}(\mathbf{m}, \sigma^2 \mathbf{C})
\end{equation}
ensuring that the search focuses around the current best estimate $\mathbf{m}$, guided by the covariance matrix $\mathbf{C}$'s shape. The CMA-ES algorithm steps are illustrated in Fig. \ref{fig:CMA_ES_principle}.

This algorithm is very efficient to find a global minimum of the minimization function, still this function must reflect the actual inversion problem and is discussed in section \ref{sec3}.

\section{Minimization function}
\label{sec3}
The search is performed in N dimensions corresponding to the number of input variables. In this case, it corresponds to the number of particles inside each energy bin. As it can involve several orders of magnitude, we perform a logarithmic transform to reduce the search scale. To estimate the fitness function a comprehensive understanding of two critical variables is essential: the experimental data ($D$) and the simulated energy deposition ($S$). The experimental data corresponds to the measure of the energy deposition inside each layer of the stack, whether from the Imaging Plate (IP) value, from the scintillator emission collected by a camera or by a differential filtering placed in front of a X-ray camera. The simulation output corresponds to the simulation of the energy deposition inside the detector by a Monte-Carlo simulation. In this section we discuss the creation of this minimization function so that it correctly transcribes the inversion problem we are dealing with and actions taken to speed up the convergence of the machine learning algorithm.

\subsection{Response Matrix}
\label{section:Response_matrix}

The deployment of Monte-Carlo simulations at each calculation of the fitness function is too computationally intensive. Each function evaluation within the CMA-ES algorithm, if reliant on direct calls to the Monte-Carlo code, could cumulatively demand several thousand CPU-hours. This high computational cost is primarily due to the intricate physics involved in the interactions of photons with the materials in the scintillators, as well as the need to accurately simulate the secondary particle generation and their subsequent interactions.

We can greatly reduce the computing time using the approximation that incident photons do not interact with the secondary particles generated within the detectors. This assumption allows us to treat the energy deposition from a continuous photon distribution as a sum of energy depositions from individual mono-energetic photons. 

The concept of the Response Matrix (RM) \cite{hannasch_compact_2021, istokskaia_experimental_2021} is as a crucial tool in this simplified model. The RM encapsulates the relationship between the energy of mono-energetic photons and the resultant energy deposition in the detector. This relationship links the hypothesized photon distribution to the observed energy deposition allowing us to define the simulated energy deposited $S$ of the scintillator $j$ as :

\begin{equation}
S_j = \sum\limits_{i=1}^N R_{i,j} x_i
\label{eq:fitness_function_sum}
\end{equation}

Where $R_{i,j}$ represents the response of scintillator/layer $j$ to mono-energetic photons at energy $E_i$ and $x_i$ signifies the value of the photon distribution at that specific energy. The sum over $N$ elements reflects the discretization of the continuous photon energy spectrum.

\subsection{Residuals}
\label{sub1sec3}

Using the RM as a speed-up process, we can now identify the key factors that must be considered in the minimization function. The first and most evident parameter is the difference between the simulated deposited energy, $S$, and the experimental data, $D$. This difference is squared to ensure positivity and to strongly amplify deviations from the experimental data. To maintain consistency across different orders of magnitude in the measured experimental data and to account for them in the same way as higher values, we normalize this difference by dividing it by the squared experimental data.
We obtain this way the weight $w_{simple}$ for the fitness function $f$ to minimize :

\begin{equation}
w_{simple} = \sum_{j}\left(\frac{D_j - S_j}{D_j}\right)^2 = \sum_{j}\left(\frac{D_j - \sum\limits_{i=1}^N R_{i,j} x_i}{D_j}\right)^2
\label{eq:fitness_simple}
\end{equation}

This simple weight can be difficult to deal with when $S_j$ is close to $D_j$. Indeed, a small discrepancy in the RM calculation can lead to convergence towards false solutions. To overcome this issue, we use a pseudo Huber loss \cite{wang_object_2019,huang_robust_2021} transforming this square behavior to a linear one for small residuals values. The weight for the difference between experimental data and simulated energy deposition $w_d$ is defined as :

\begin{equation}
\begin{split}
w_{d} 
&= \delta^2\left(\sqrt{1+\left(\frac{w_{\mathrm{simple}}}{\delta}\right)^2}-1\right) \\
&= \delta^2\left(
    \sqrt{1+\sum_{j}\left(
        \frac{D_j - \sum\limits_{i=1}^N R_{i,j} x_i}{\delta D_j}
    \right)^2}
    -1
\right)
\label{eq:weight_calibration_factor}
\end{split}
\end{equation}

\begin{figure}[H]
\centering
\includegraphics[width = 0.48\textwidth]{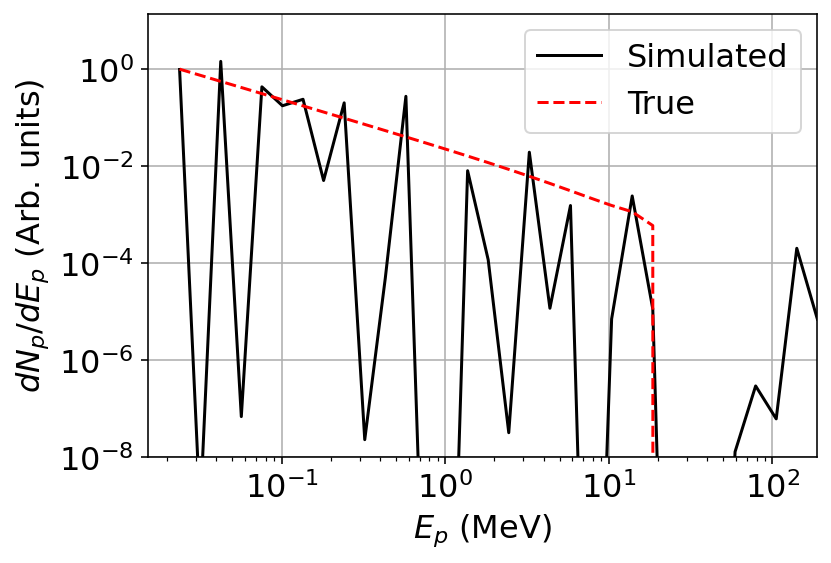}
\caption{Unfolding of a numerically generated Bremsstrahlung distribution leading to "peak" solutions.}
\label{fig:Unfolding_peak}
\end{figure}

Using this minimization function tends to converge towards peak solutions where only specific energies are set to the correct value. This works well for mono-energetic sources such as radioactive sources but fails to grasp continuous distribution due to this under-determined system where more energy bins need to be unfolded than are detectors layers as shown in Fig. \ref{fig:Unfolding_peak}.

To remove this peak solution, a more robust minimization function must be created to overcome the under-determinity of the system and obtain a smoother spectrum distribution.

\subsection{Smoothing factor}
\label{sec:Smoothing_factor}

The convergence towards peaks solutions can be overcome by adding a weight that takes into consideration this continuity. As the search is performed in log-space, trying to directly minimize the first order derivative of the spectrum will introduce a bias towards pure power-law spectrum, ie Bremsstrahlung. This is why, we instead try to minimize the second order derivative which allow for more freedom around this power-law. This smoothing factor $w^{(2)}$ is defined using a second order derivative :
\begin{equation}
w^{(2)} = a_s\sum_{i=1}^{N-2} 
\left(x_{i+2} - 2 x_{i+1} + x_i\right)^2
\label{eq:smoothing_second_order}
\end{equation}

Using $a_s$ a weight factor so that $w^{(2)}$ and $w_{d}$ are comparable and $N$ the number of points in the energy search space. This weight works well for most applications, however it tends to smooth out hard cut-off as it can be observed for bremsstrahlung or synchrotron radiation as shown in Fig. \ref{fig:Unfolding_bremsstrahlung} and Fig. \ref{fig:Unfolding_synchrotron}. To compensate this effect, it is possible to apply an adaptive smoothing factor. It takes the shape of a stronger weight for lower energies than at higher energies using a sigmoid function as factor, giving for this adaptive weight $w_a^{(2)}$:

\begin{equation}
w_a^{(2)} = \sum_{i=1}^{N-2} 
\left(x_{i+2} - 2 x_{i+1} + x_i\right)^2w_i^{sg}
\label{eq:adapt_weighting}
\end{equation}
\begin{equation}
w_i^{sg} = \frac{1}{1+e^{\sigma_{sg}(i-i_{start})}}
\label{eq:bin_smoothing}
\end{equation}
with $\sigma_{sg}$ the steepness of the sigmoid function, and $i_{start}$ the start of the function drop. This adaptive weight must be used carefully as it can under-smooth some parts of the spectrum and go back to the "peak" solutions. At the moment, these parameters need to be manually set and will be upgraded in the future to be included in an automatic way.

\subsection{Calibration factors}
\label{sec:Calibration_factors}

The calculation of the RM is a critical aspect of this method, as it must closely match the actual experimental setup. To mitigate errors arising from inaccuracies in RM calculations or experimental noise, the energy deposition simulation data \( S_j \) can be introduced as variables rather than being strictly determined by the RM. This is implemented by applying scaling factors \( f_j \) to the elements \( R_j \) of the RM. However, to maintain the dominance of RM-based calculations in determining \( S_j \), a gated weight function is introduced. This function strongly increases when values deviate beyond a specified range, ensuring that the \( f_j \) factors only introduce a limited modification, typically within 5\% of the originally computed RM. The gated weight \( w_f \) is defined as the sum of two soft-plus functions and a threshold percentage \( p \), effectively discouraging substantial deviations from the RM. This weight is then put to the power of 4 to enhance the slope outside of the bound while having a flat shape inside. A factor $a_f$ is used for having similar weight compared to other contributions.

\begin{equation}
w_{f} = a_f(\frac{1}{k} \sum \log(1 + e^{k (1-p - f_j)}) + \log(1 + e^{k(f_j-1 + p)}))^4 
\label{eq:smoothing}
\end{equation}

Typical values used are $k$ = 300, $p$ = 0.05 and $a_f$ = 10.

\label{sub3sec3}

\section{Computer generated data}
\label{sec4}
To assess the accuracy of the unfolding, we use known theoretical formula of three expected distributions in ultra-high intensity laser plasma interaction and ICF; bremsstrahlung, synchrotron and gaussian. The response of the detector from this spectrum is then calculated. This response is taken as the measured data for the unfolding, trying to find back the true or original spectrum.

\subsection{Unfolding from known distributions}
\label{sub2sec4}

Synchrotron-type spectra naturally emerge in ultra-relativistic laser–plasma interactions, particularly as high-energy cut-off tails. Such distributions present a demanding test for unfolding algorithms due to their extended high-energy tails and relatively low photon statistics at lower energies.

\begin{figure}[H]
\centering
\includegraphics[width = 0.4\textwidth]{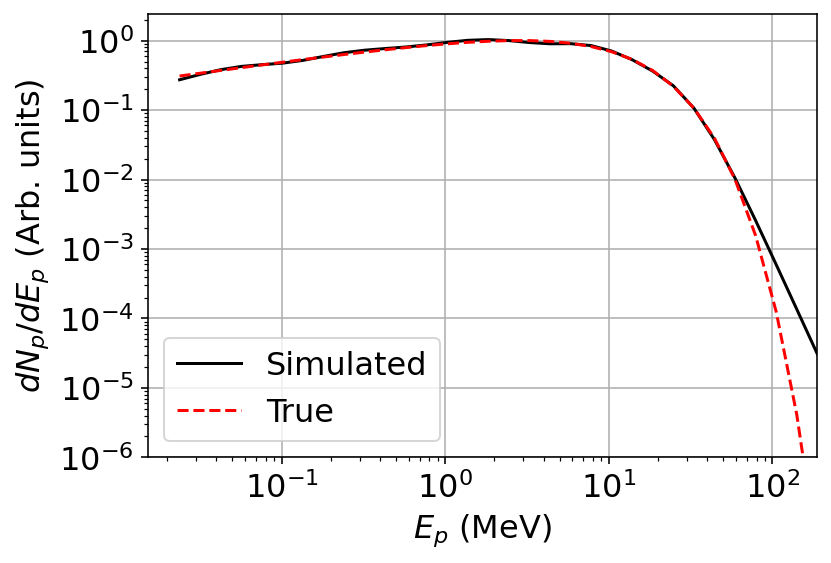}
\caption{Unfolding of a numerically generated synchrotron distribution.}
\label{fig:Unfolding_synchrotron}
\end{figure}

\begin{figure}[H]
\centering
\includegraphics[width = 0.4\textwidth]{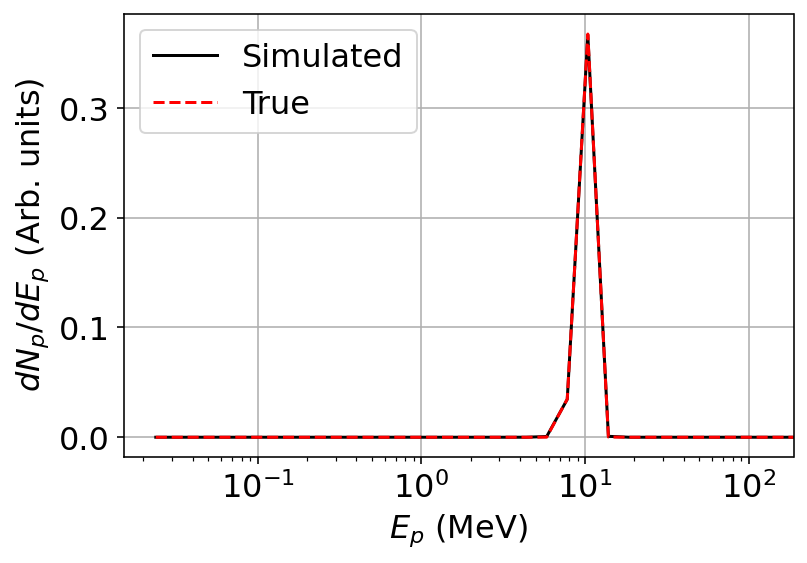}
\caption{Unfolding of a numerically generated Gaussian distribution of mean 10 MeV and standard deviation 1 MeV.}
\label{fig:Unfolding_gaussian}
\end{figure}

Figure \ref{fig:Unfolding_synchrotron} shows the reconstruction obtained by unfolding a numerically generated synchrotron spectrum. The unfolded distribution closely follows the true curve over several orders of magnitude, reproducing the characteristic rise and exponential-like decay from tens of keV to hundreds of MeV.

A similar test is performed for a narrow Gaussian spectrum, representative of scenarios where quasi-monochromatic emission is produced. Although less common in bulk bremsstrahlung from dense plasmas, Gaussian-like features are relevant for specific mechanisms such as high-order harmonic generation in solids or narrowband line emission such as calibration radioactive sources. The reconstruction, shown in Fig. \ref{fig:Unfolding_gaussian}, demonstrates that the algorithm is capable of recovering both the centroid and the width of the distribution.

\begin{figure}[H]
\centering
\includegraphics[width = 0.4\textwidth]{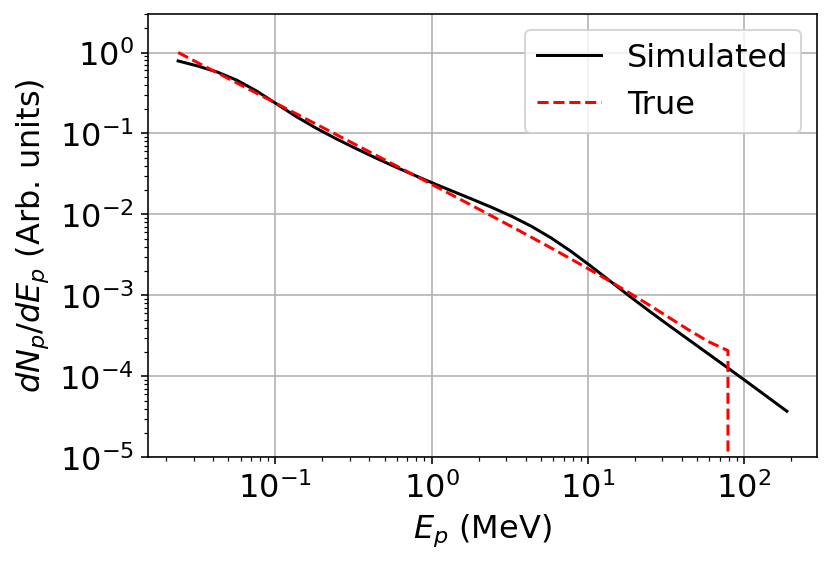}
\caption{Unfolding of a numerically generated bremsstrahlung distribution.}
\label{fig:Unfolding_bremsstrahlung}
\end{figure}

A more common distribution observed in laser–solid interactions is Bremsstrahlung signal. In many practical conditions the photon number follows an approximate power-law dependence close to an inverse-energy scaling, leading to a higher photon flux at lower energies and a smoother energy deposition across the detector layers. This makes bremsstrahlung spectra particularly favourable test cases for unfolding, as the signal amplitude is well distributed and the dynamic range is naturally compatible with the detector response. Figure \ref{fig:Unfolding_bremsstrahlung} compares the true and reconstructed bremsstrahlung spectra for a synthetic test distribution. The excellent agreement between the two curves confirms that the unfolding procedure performs robustly for this class of spectra. 

Across these examples, the reconstructed spectra highlight both the strengths and the natural limitations of unfolding from stacked calorimeters. In the synchrotron case, the method reliably captures photons from approximately $50$ keV up to $100$ MeV. At the lowest energies, however, the detector sensitivity decreases, leading to weaker signals that may be partially overshadowed by contributions from higher-energy photons. This imbalance results in a bias in the reconstructed distribution, particularly visible when the deposited energy becomes small relative to that of higher-energy particles. Conversely, at the highest energies, above roughly $90$ MeV in this test case, the unfolded spectrum exhibits a smoother roll-off than the sharp cut-off in the true distribution. This behaviour arises from the interplay between the smoothing regularization and the finite precision with which high-energy deposition profiles can be distinguished. An adaptive smoothing scheme, discussed in Sections \ref{sec:Smoothing_factor} and \ref{sec:Smoothing_factor_data}, mitigates this effect by adjusting the regularization strength dynamically across the spectrum.

\subsection{Double type of particles}
\label{sub3sec4}

In realistic experimental environments, detector signals may include contributions from multiple particle species. High-energy electrons, for example, can traverse magnetic or shielding structures and strike the detector directly, superimposing their energy deposition on the photon-induced bremsstrahlung response. To address such cases, the unfolding framework can be extended to include additional spectral components, each represented by its own set of parameters within the optimization space. This allows the algorithm to simultaneously reconstruct the spectra of different populations, provided that their respective response functions are sufficiently distinct.
\begin{figure}[H]
    \centering
    \begin{minipage}{0.4\textwidth}
        \centering
        \includegraphics[width=\linewidth]{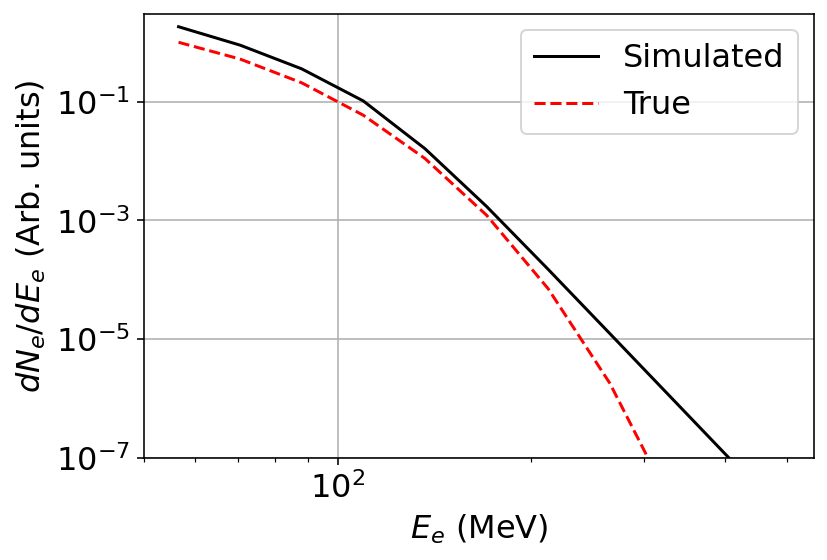}
        \caption{Unfolding of a Maxwell–Jüttner electron distribution of temperature $25$~MeV alongside a bremsstrahlung photon signal showed in Fig. \ref{fig:graph2}. Low energies electrons are not included to replicate the effect of magnets typically placed before the detector.}
        \label{fig:graph1}
    \end{minipage}\hfill
    \begin{minipage}{0.4\textwidth}
        \centering
        \includegraphics[width=\linewidth]{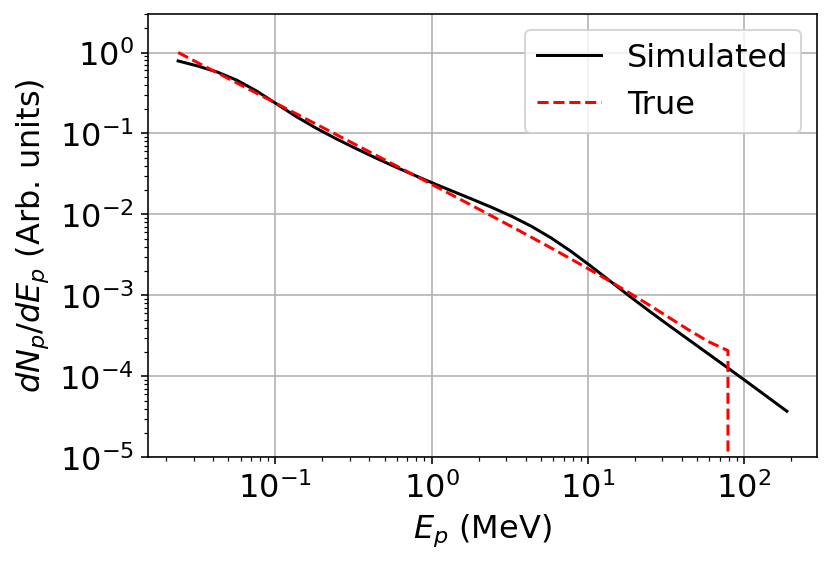}
        \caption{Unfolding of a bremsstrahlung photon distribution with an added Maxwell–Jüttner electron contribution showed in Fig. \ref{fig:graph1}.}
        \label{fig:graph2}
    \end{minipage}
\end{figure}

To illustrate this capability, we generate synthetic measurements consisting of a bremsstrahlung photon spectrum combined with a Maxwell–Jüttner electron distribution of temperature $25$ MeV representing a total of 75\% of the total energy deposited inside the detector. The unfolding is performed with both components active, enabling the algorithm to separate the photon and electron contributions. Figures \ref{fig:graph1} and \ref{fig:graph2} show the reconstructed spectra for each species. The method succeeds in disentangling the two populations and recovers both spectral shapes with good fidelity, demonstrating its applicability to experimental scenarios where mixed particle fluxes are unavoidable.

\subsection{Noisy unfolding}
\label{sub3sec5}

In many experimental configurations, the measured depth-dose profile is unavoidably affected by noise. Such perturbations can originate from a variety of sources, including parasitic radiation generated by secondary interactions, high-energy particles striking the detector directly, fluctuations in the detector response due to environmental conditions, or discrepancies between the idealized response matrix and the actual experimental geometry. Because stacked calorimeters integrate signals over multiple layers, even small layer-dependent deviations can accumulate and distort the reconstructed spectrum if not accounted for. It is therefore essential for any unfolding framework to incorporate mechanisms that mitigate these effects while preserving the stability and fidelity of the reconstruction.
\begin{figure}[H]
\centering
\includegraphics[width = 0.4\textwidth]{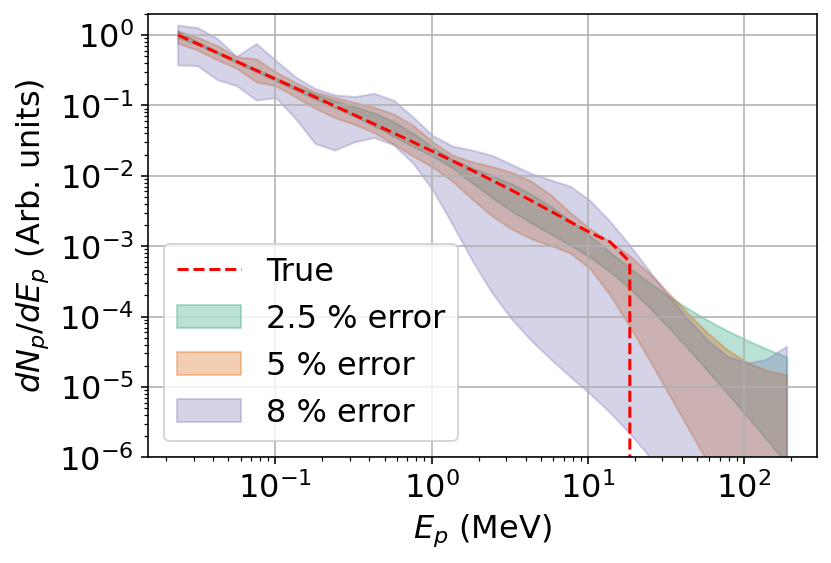}
\caption{Unfolding of a numerically generated bremsstrahlung distribution with added layer-dependent noise. The dashed red curve represents the true distribution, while the colored bands denote the standard deviation obtained from 30 independent unfoldings. For each unfolding, random noise was added to every layer, sampled from a uniform distribution with the width indicated in the corresponding label.}
\label{fig:Unfolding_noise}
\end{figure}
To address this, the algorithm introduces an additional set of optimization variables representing layer-dependent calibration factors. These factors allow each scintillator or detector layer to deviate slightly from its nominal response, typically within a few percent, thereby absorbing a portion of the noise and compensating for systematic or quasi-systematic discrepancies. As discussed in Section \ref{sec:Calibration_factors}, this approach provides a flexible means of correcting for layer-specific drifts or imperfections without imposing strong assumptions on the spectral shape itself. By constraining the allowed variation to a narrow interval, the method avoids unphysical distortions while improving the agreement between the measured and modelled signals.

Figure \ref{fig:Unfolding_noise} illustrates the impact of this procedure using a synthetic bremsstrahlung distribution to which random noise has been added independently on each scintillator. Each colored band represents the standard deviation over 30 independent unfoldings, corresponding to a different total deviation. The value indicated in the label specifies the bounds of the uniform distribution from which the noise was sampled. Despite the significant layer-to-layer fluctuations, the unfolding consistently reconstructs the global spectral trend, up to at least 5\% demonstrating that the calibration-factor extension effectively stabilizes the inversion. Local deviations remain visible in regions where the detector is intrinsically less sensitive, but the ability to recover the overall spectral slope and cut-off energy shows that the reconstruction remains robust even under substantial noise conditions. This capability is particularly valuable in high-power laser environments, where shot-to-shot variability and secondary particle contamination are common and cannot always be suppressed at the hardware level. Note that this level of noise corresponds to a limit-test and no experimental results should reach this level.

\subsection{Adaptive smoothing factor}
\label{sec:Smoothing_factor_data}

\begin{figure}[H]
\centering
\includegraphics[width = 0.4\textwidth]{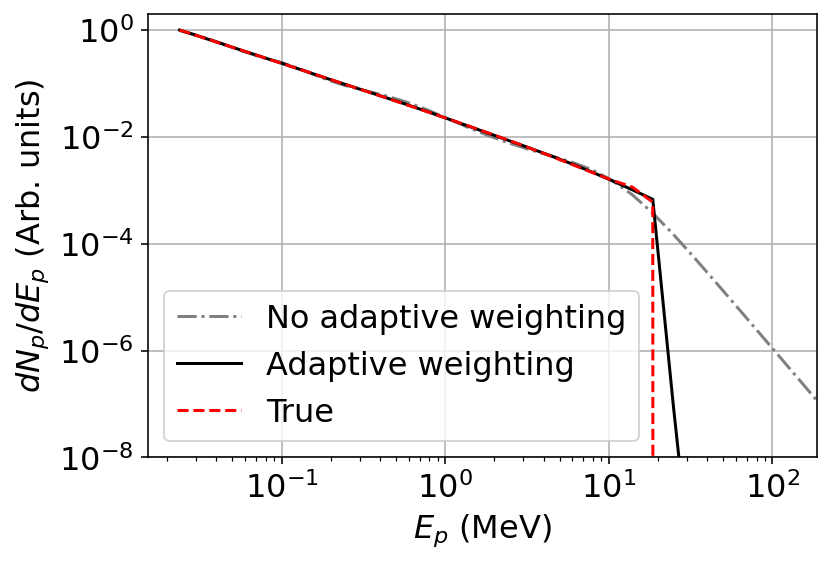}
\caption{Unfolding of a Bremsstrahlung distribution using the typical smoothing factor, labeled as no adaptive weighting, or with the adaptive smoothing factor described in Fig. \ref{eq:adapt_weighting}.}
\label{fig:Adaptive_factor}
\end{figure}

As shown in Fig. \ref{fig:Unfolding_synchrotron}, a sharp cut-off in the energy spectrum can be difficult to reproduce accurately, because the smoothing factor may dominate over the term representing the difference between the simulated response and the measured data. To compensate for this effect, it is possible to reduce the value of the smoothing weight near the cut-off as discussed in \ref{sec:Smoothing_factor}. By doing so it is possible to better catch the cut-off as showed in Fig. \ref{fig:Adaptive_factor}.
\subsection{Typical computing time}
\label{sec:Time}
State-of-the-art detectors start to go towards the real-time analysis of the data. To do so, the unfolding algorithm must be fast and efficient as well. In the table hereafter we review the different processing time, number of function evaluation and the impact of a good first guess, taken as the true spectrum and an added noise in the percent range to it, refered to as First Guess Reduction Time (FGRT) expressed in percent corresponding to the reduced time compared to a random first guess. Each calculated for different cases using 32 bins in the spectrum run locally on lab-grade computer. The algorithm used here has a $N\log(N)$ computing time dependency for the simplest case. Results are summarized in Table \ref{tab:CMA_time}.

\begin{table}[H]
    \centering
    \begin{tabular}{|c|c|c|c|}
       \hline
       Case   & Time (s) & Function evaluations  & FGRT (\%)\\ \hline
        Simple &   37      &         4.2e6              &      32     \\   \hline
       Adaptive weighting (AW)&     51     &   4.6e6    &      30    \\  \hline
       Calibration factors (CF)&   87     &   5.7e6      &     0   \\ \hline
       AW + CF &   123     &    6.7e6     &    0    \\ \hline
       
    \end{tabular}
    \caption{Computing time for different weights used along with the number of function evaluations. The impact of a first good guess on the reduction on the computing time (FGRT) is also included.}
    \label{tab:CMA_time}
\end{table}
\section{Detector designed at ELI-Beamlines and experimental measure}
\label{sub1sec5}

\begin{figure}[H]
\centering

    \includegraphics[width=\linewidth]{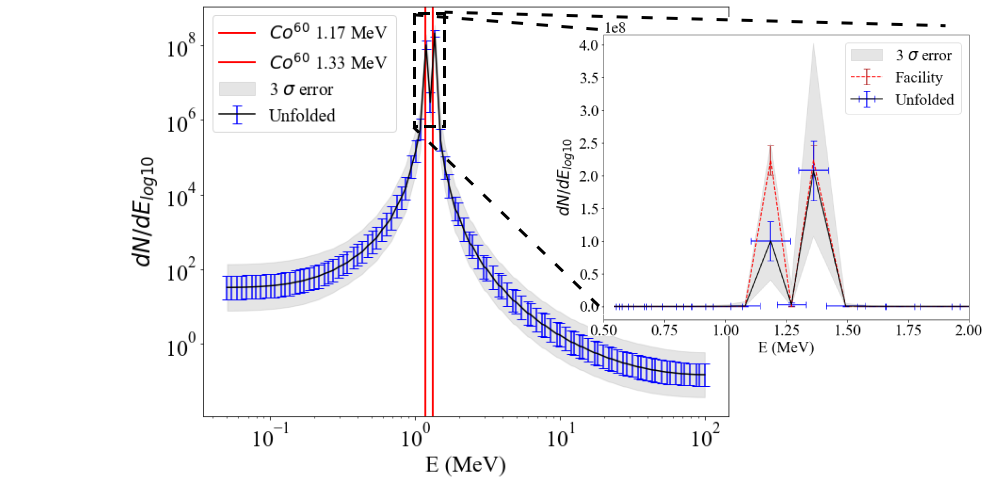}
    \caption{Unfolded spectrum from a Co$^{60}$ radioactive source.}
    \label{fig:Co60_calibration}
\end{figure}

A detector using stacking scintillators have been designed at ELI-Beamlines described in reference \cite{fauvel_compact_2025}. It is commonly used inside vacuum in order to reduce the Bremsstrahlung signal coming from high energy electrons inside the flange and also to maximize the photon flux on the detector, obtaining higher signal to noise ratio. To determine the calibration factor of each scintillator and assess the detector energy resolution, a calibration using a radioactive Co$^{60}$ source was used. This source possesses two closely separated photopeaks, ideal for energy resolution assessment. Fig. \ref{fig:Co60_calibration} shows the unfolded spectrum compared to the values provided by the calibration facility.

The unfolded data is in extremely good agreement with the facility data, showing the accuracy of the unfolding algorithm being able to resolve such closely separated photopeaks and also being able to reproduce the photon number with a good accuracy. Besides, a second spectrometer using a different set of scintillators was also calibrated and gave similar results showcasing the repeatability of the algorithm.

\section{Conclusion}

We have presented an open-source unfolding framework tailored for stacking–calorimeter diagnostics, with particular emphasis on applications relevant to high-energy-density physics, the inertial confinement fusion community and ultra-high intensity community. The algorithm demonstrates high spectral fidelity across a broad range of conditions, recovering closely spaced $\gamma$-ray photopeaks, such as those from a \textsuperscript{60}Co calibration source, with accurate photon-yield reconstruction. Its robustness to substantial layer dependent noise with up to 5 percent deviation in the tests presented here shows that the method remains reliable even under the challenging conditions typical of high-intensity laser–plasma experiments, where secondary radiation, target debris, and detector mismatch frequently distort raw measurements.

For ICF facilities, where bremsstrahlung cannons and related stacked diagnostics serve as essential tools for quantifying hot-electron populations, preheat levels, and hard X-ray emission, the unfolding capability introduced here provides a direct means of improving the interpretability and fidelity of these measurements. By avoiding assumptions about the spectral shape and by leveraging a noise-resilient optimization strategy, the approach supports precise reconstruction of bremsstrahlung spectra over large dynamic ranges, enabling more accurate assessments of suprathermal electron transport and energy deposition, quantities that critically influence implosion symmetry, fuel adiabat, and overall target performance. The algorithm therefore complements ongoing diagnostic developments at major laser facilities and provides a transferable, facility-independent method compatible with both passive and active stacking detectors.

\section*{Acknowledgment}
We wish to acknowledge the support of the National Sci-
ence Foundation (NSF Grant No. PHY-2206777) and the Czech Science Foundation
(GA ČR) for funding on project number No. 22-42890L in the frame of the National Science Foundation–Czech Science Foundation partnership.

\section*{Data Availability Statement}

The data that support the findings of this study are available from the corresponding author upon reasonable request.
\section*{Data Availability Statement}
The authors have no conflicts to disclose.
\section*{Bibliography}
\bibliography{aipsamp}

@misc{nikolaus_hansen_cma_2016,
	title = {The {CMA} {Evolution} {Strategy}: {A} {Tutorial}},
	shorttitle = {The {CMA} {Evolution} {Strategy}},
	url = {http://arxiv.org/abs/1604.00772},
	urldate = {2024-07-03},
	publisher = {arXiv},
	author = {Nikolaus Hansen},
	year = {2016},
	note = {arXiv:1604.00772 [cs, stat]},
}

@misc{hansen_cma-espycma_2024,
	title = {{CMA}-{ES}/pycma: r3.4.0},
	shorttitle = {{CMA}-{ES}/pycma},
	url = {https://zenodo.org/records/13097780},
	urldate = {2024-08-26},
	publisher = {Zenodo},
	author = {Hansen, Nikolaus and yoshihikoueno and ARF1 and Cakmak, Sait and Kadlecová, Gabi and Nozawa, Kento and Rolshoven, Luca and Akimoto, Youhei and brieglhostis and Brockhoff, Dimo},
	month = jul,
	year = {2024},
	doi = {10.5281/zenodo.13097780},
}

@article{hannasch_compact_2021,
	title = {Compact spectroscopy of {keV} to {MeV} {X}-rays from a laser wakefield accelerator},
	volume = {11},
	issn = {2045-2322},
	url = {https://www.nature.com/articles/s41598-021-93689-5},
	doi = {10.1038/s41598-021-93689-5},
	language = {en},
	number = {1},
	urldate = {2022-09-09},
	journal = {Scientific Reports},
	author = {Hannasch, A. and Laso Garcia, A. and LaBerge, M. and Zgadzaj, R. and Köhler, A. and Couperus Cabadağ, J. P. and Zarini, O. and Kurz, T. and Ferrari, A. and Molodtsova, M. and Naumann, L. and Cowan, T. E. and Schramm, U. and Irman, A. and Downer, M. C.},
	month = dec,
	year = {2021},
	pages = {14368},
}

@article{istokskaia_experimental_2021,
	title = {Experimental tests and signal unfolding of a scintillator calorimeter for laser-plasma characterization},
	volume = {16},
	issn = {1748-0221},
	url = {https://dx.doi.org/10.1088/1748-0221/16/02/T02006},
	doi = {10.1088/1748-0221/16/02/T02006},
	language = {en},
	number = {02},
	urldate = {2023-08-01},
	journal = {Journal of Instrumentation},
	author = {Istokskaia, V. and Stránský, V. and Giuffrida, L. and Versaci, R. and Grepl, F. and Tryus, M. and Velyhan, A. and Dudžák, R. and Krása, J. and Krupka, M. and Singh, S. and Neely, D. and Olšovcová, V. and Margarone, D.},
	month = feb,
	year = {2021},
	pages = {T02006},
}

@article{rhee_spectral_2016,
	title = {Spectral tomographic analysis of {Bremsstrahlung} {X}-rays generated in a laser-produced plasma},
	volume = {34},
	issn = {0263-0346, 1469-803X},
	url = {https://www.cambridge.org/core/journals/laser-and-particle-beams/article/spectral-tomographic-analysis-of-bremsstrahlung-xrays-generated-in-a-laserproduced-plasma/0F10E009450AABCC2D3F1FC0072F17E6},
	doi = {10.1017/S0263034616000604},
	language = {en},
	number = {4},
	urldate = {2023-07-28},
	journal = {Laser and Particle Beams},
	author = {Rhee, Y. J. and Nam, S. M. and Peebles, J. and Sawada, H. and Wei, M. and Vaisseau, X. and Sasaki, T. and Giuffrida, L. and Hulin, S. and Vauzour, B. and Santos, J. J. and Batani, D. and McLean, H. S. and Patel, P. K. and Li, Y. T. and Yuan, D. W. and Zhang, K. and Zhong, J. Y. and Fu, C. B. and Hua, N. and Li, K. and Zhang, Y. and Zhu, J. Q. and Kim, I. J. and Jeon, J. H. and Jeong, T. M. and Choi, I. W. and Lee, H. W. and Sung, J. H. and Lee, S. K. and Nam, C. H.},
	month = dec,
	year = {2016},
	note = {Publisher: Cambridge University Press},
	pages = {645--654},
}

@article{fauvel_compact_2025,
	title = {Compact in-vacuum gamma-ray spectrometer for high-repetition rate {PW}-class laser–matter interaction},
	volume = {96},
	url = {https://pubs.aip.org/aip/rsi/article/96/2/023102/3334207},
	number = {2},
	urldate = {2025-04-21},
	journal = {Review of Scientific Instruments},
	author = {Fauvel, G. and Tangtartharakul, K. and Arefiev, A. and De Chant, J. and Hakimi, S. and Klimo, O. and Manuel, M. and McIlvenny, A. and Nakamura, K. and Obst-Huebl, L.},
	year = {2025},
	note = {Publisher: AIP Publishing},
}

@article{tavana_ultra-high_2023,
	title = {Ultra-high efficiency bremsstrahlung production in the interaction of direct laser-accelerated electrons with high-{Z} material},
	volume = {11},
	issn = {2296-424X},
	url = {https://www.frontiersin.org/journals/physics/articles/10.3389/fphy.2023.1178967/full},
	doi = {10.3389/fphy.2023.1178967},
	language = {English},
	urldate = {2025-11-20},
	journal = {Frontiers in Physics},
	author = {Tavana, P. and Bukharskii, N. and Gyrdymov, M. and Spillmann, U. and Zähter, Ş and Cikhardt, J. and Borisenko, N. G. and Korneev, Ph and Jacoby, J. and Spielmann, C. and Andreev, N. E. and Günther, M. M. and Rosmej, O. N.},
	month = may,
	year = {2023},
	note = {Publisher: Frontiers},
}

@misc{noauthor_user_2024,
	title = {User {Guide} {LMJ} {Laser} {MegaJoule} {PETAL} {PETawatt} {Aquitaine} {Laser}},
	url = {https://www-lmj.cea.fr/docs/2024/LMJ-PETAL-Users-guide-v2.1.pdf},
	urldate = {2025-11-20},
	year = {2024},
}

@article{koester_bremsstrahlung_2021,
	title = {Bremsstrahlung cannon design for shock ignition relevant regime},
	volume = {92},
	issn = {0034-6748},
	url = {https://doi.org/10.1063/5.0022030},
	doi = {10.1063/5.0022030},
	number = {1},
	urldate = {2025-11-20},
	journal = {Review of Scientific Instruments},
	author = {Koester, P. and Baffigi, F. and Cristoforetti, G. and Labate, L. and Gizzi, L. A. and Baton, S. and Koenig, M. and Colaïtis, A. and Batani, D. and Casner, A. and Raffestin, D. and Tentori, A. and Trela, J. and Rousseaux, C. and Boutoux, G. and Brygoo, S. and Jacquet, L. and Reverdin, C. and Le Bel, E. and Le-Deroff, L. and Theobald, W. and Shigemori, K.},
	month = jan,
	year = {2021},
	pages = {013501},
}

@phdthesis{kirby_radiation_2011,
	type = {d\_ph},
	title = {Radiation dosimetry of conventional and laser-driven particle beams},
	url = {https://etheses.bham.ac.uk/id/eprint/2816/},
	language = {English},
	urldate = {2025-11-20},
	school = {University of Birmingham},
	author = {Kirby, Daniel James},
	month = dec,
	year = {2011},
}

@article{nurnberg_radiochromic_2009,
	title = {Radiochromic film imaging spectroscopy of laser-accelerated proton beams},
	volume = {80},
	issn = {0034-6748},
	url = {https://doi.org/10.1063/1.3086424},
	doi = {10.1063/1.3086424},
	number = {3},
	urldate = {2025-11-20},
	journal = {Review of Scientific Instruments},
	author = {Nürnberg, F. and Schollmeier, M. and Brambrink, E. and Blažević, A. and Carroll, D. C. and Flippo, K. and Gautier, D. C. and Geißel, M. and Harres, K. and Hegelich, B. M. and Lundh, O. and Markey, K. and McKenna, P. and Neely, D. and Schreiber, J. and Roth, M.},
	month = mar,
	year = {2009},
	pages = {033301},
}

@article{schmitz_automated_2022,
	title = {Automated reconstruction of the initial distribution of laser accelerated ion beams from radiochromic film ({RCF}) stacks},
	volume = {93},
	issn = {0034-6748},
	url = {https://doi.org/10.1063/5.0094105},
	doi = {10.1063/5.0094105},
	number = {9},
	urldate = {2025-11-20},
	journal = {Review of Scientific Instruments},
	author = {Schmitz, Benedikt and Metternich, Martin and Boine-Frankenheim, Oliver},
	month = sep,
	year = {2022},
	pages = {093306},
}

@article{beyer_toward_2017,
	title = {Toward a {Steady}-{State} {Analysis} of an {Evolution} {Strategy} on a {Robust} {Optimization} {Problem} {With} {Noise}-{Induced} {Multimodality}},
	volume = {21},
	issn = {1941-0026},
	url = {https://ieeexplore.ieee.org/abstract/document/7850990},
	doi = {10.1109/TEVC.2017.2668068},
	number = {4},
	urldate = {2025-11-20},
	journal = {IEEE Transactions on Evolutionary Computation},
	author = {Beyer, Hans-Georg and Sendhoff, Bernhard},
	month = aug,
	year = {2017},
	pages = {629--643},
}

@article{craveiro_unfolding_2024,
	title = {Unfolding experimental distortions in beta spectrometry},
	volume = {12},
	issn = {2296-424X},
	url = {https://www.frontiersin.org/journals/physics/articles/10.3389/fphy.2024.1435615/full},
	doi = {10.3389/fphy.2024.1435615},
	language = {English},
	urldate = {2025-11-20},
	journal = {Frontiers in Physics},
	author = {Craveiro, Gaël and Leblond, Sylvain and Mougeot, Xavier and Vivier, Matthieu},
	month = aug,
	year = {2024},
	note = {Publisher: Frontiers},
}

@article{reginatto_spectrum_2002,
	series = {Int. {Workshop} on {Neutron} {Field} {Spectrometry} in {Science}, {Technolog} y and {Radiation} {Protection}},
	title = {Spectrum unfolding, sensitivity analysis and propagation of uncertainties with the maximum entropy deconvolution code {MAXED}},
	volume = {476},
	issn = {0168-9002},
	url = {https://www.sciencedirect.com/science/article/pii/S0168900201014395},
	doi = {10.1016/S0168-9002(01)01439-5},
	number = {1},
	urldate = {2025-11-20},
	journal = {Nuclear Instruments and Methods in Physics Research Section A: Accelerators, Spectrometers, Detectors and Associated Equipment},
	author = {Reginatto, Marcel and Goldhagen, Paul and Neumann, Sonja},
	month = jan,
	year = {2002},
	pages = {242--246},
}

@article{istokskaia_real-time_2024,
	title = {Real-time bremsstrahlung detector as a monitoring tool for laser–plasma proton acceleration},
	volume = {12},
	issn = {2095-4719, 2052-3289},
	url = {https://www.cambridge.org/core/journals/high-power-laser-science-and-engineering/article/realtime-bremsstrahlung-detector-as-a-monitoring-tool-for-laserplasma-proton-acceleration/A281847E8274B1F042AD3F9C72563ADC},
	doi = {10.1017/hpl.2024.38},
	language = {fr},
	urldate = {2025-11-20},
	journal = {High Power Laser Science and Engineering},
	author = {Istokskaia, Valeria and Lefebvre, Benoit and Versaci, Roberto and Dreghici, Dragana B. and Doria, Domenico and Grepl, Filip and Olšovcová, Veronika and Schillaci, Francesco and Stanček, Stanislav and Tryus, Maksym and Velyhan, Andriy and Margarone, Daniele and Giuffrida, Lorenzo},
	month = jan,
	year = {2024},
	pages = {e57},
}

@article{breschi_new_2004,
	title = {A new algorithm for spectral and spatial reconstruction of proton beams from dosimetric measurements},
	volume = {522},
	issn = {0168-9002},
	url = {https://www.sciencedirect.com/science/article/pii/S0168900203031449},
	doi = {10.1016/j.nima.2003.11.199},
	number = {3},
	urldate = {2025-11-20},
	journal = {Nuclear Instruments and Methods in Physics Research Section A: Accelerators, Spectrometers, Detectors and Associated Equipment},
	author = {Breschi, E. and Borghesi, M. and Galimberti, M. and Giulietti, D. and Gizzi, L. A. and Romagnani, L.},
	month = apr,
	year = {2004},
	pages = {190--195},
}

@article{seimetz_spectral_2018,
	title = {Spectral characterization of laser-accelerated protons with {CR}-39 nuclear track detector},
	volume = {89},
	issn = {0034-6748},
	url = {https://doi.org/10.1063/1.5009587},
	doi = {10.1063/1.5009587},
	number = {2},
	urldate = {2025-11-20},
	journal = {Review of Scientific Instruments},
	author = {Seimetz, M. and Bellido, P. and García, P. and Mur, P. and Iborra, A. and Soriano, A. and Hülber, T. and García López, J. and Jiménez-Ramos, M. C. and Lera, R. and Ruiz-de la Cruz, A. and Sánchez, I. and Zaffino, R. and Roso, L. and Benlloch, J. M.},
	month = feb,
	year = {2018},
	pages = {023302},
}

@article{ghosh_effect_2022,
	title = {Effect of soft and hard x-rays on shock propagation, preheating and ablation characteristics in pure and doped {Be} ablators},
	volume = {29},
	issn = {1070-664X, 1089-7674},
	url = {http://arxiv.org/abs/2108.02933},
	doi = {10.1063/5.0090598},
	number = {6},
	urldate = {2025-11-21},
	journal = {Physics of Plasmas},
	author = {Ghosh, Karabi and Mishra, Gaurav},
	month = jun,
	year = {2022},
	note = {arXiv:2108.02933 [physics]},
	pages = {062703},
}

@article{chaurasia_optimization_2013,
	title = {Optimization of bremsstrahlung and characteristic line emission from aluminum plasma},
	volume = {308},
	issn = {0030-4018},
	url = {https://www.sciencedirect.com/science/article/pii/S0030401813006329},
	doi = {10.1016/j.optcom.2013.07.001},
	urldate = {2025-11-21},
	journal = {Optics Communications},
	author = {Chaurasia, S. and Tripathi, S. and Leshma, P. and Murali, C. G. and Pasley, J.},
	month = nov,
	year = {2013},
	pages = {169--174},
}

@article{tommasini_development_2011,
	title = {Development of {Compton} radiography of inertial confinement fusion implosionsa)},
	volume = {18},
	issn = {1070-664X},
	url = {https://ui.adsabs.harvard.edu/abs/2011PhPl...18e6309T},
	doi = {10.1063/1.3567499},
	urldate = {2025-11-21},
	journal = {Physics of Plasmas},
	author = {Tommasini, R. and Hatchett, S. P. and Hey, D. S. and Iglesias, C. and Izumi, N. and Koch, J. A. and Landen, O. L. and MacKinnon, A. J. and Sorce, C. and Delettrez, J. A. and Glebov, V. Yu. and Sangster, T. C. and Stoeckl, C.},
	month = may,
	year = {2011},
	note = {Publisher: AIP
ADS Bibcode: 2011PhPl...18e6309T},
	pages = {056309},
}

@article{huang_robust_2021,
	title = {Robust pairwise learning with {Huber} loss},
	volume = {66},
	issn = {0885-064X},
	url = {https://www.sciencedirect.com/science/article/pii/S0885064X2100025X},
	doi = {10.1016/j.jco.2021.101570},
	urldate = {2025-11-25},
	journal = {Journal of Complexity},
	author = {Huang, Shouyou and Wu, Qiang},
	month = oct,
	year = {2021},
	pages = {101570},
}

@article{wang_object_2019,
	title = {Object tracking based on {Huber} loss function},
	volume = {35},
	issn = {1432-2315},
	url = {https://doi.org/10.1007/s00371-018-1563-1},
	doi = {10.1007/s00371-018-1563-1},
	language = {en},
	number = {11},
	urldate = {2025-11-25},
	journal = {The Visual Computer},
	author = {Wang, Yong and Hu, Shiqiang and Wu, Shandong},
	month = nov,
	year = {2019},
	pages = {1641--1654},
}
\bibliographystyle{unsrt}
\end{document}